# On the nature of the negative-conductivity resonance in a superlattice Bloch oscillator

Herbert Kroemer[*]

ECE Department, University of California, Santa Barbara, CA 93106

## Abstract

Adding a high-frequency ac component to the bias field of a superlattice induces a synchronous modulation of the velocity with which the electrons traverse the Brillouin zone. In the presence of inelastic scattering, the *k*-space velocity modulation causes a wave-like bunching of the electrons in *k*-space, which in turn induces a high-frequency component in the real-space current, synchronous with the drive field, but phase-shifted relative to the latter. For a drive frequency $\omega$ equal to the Bloch frequency $\omega_B$, the phase shift is less than 90° (implying a positive ac conductivity), increasing to 90° in the limit of a vanishing scattering (a purely reactive current). If the drive frequency is lowered below the Bloch frequency, the phase shift can increase beyond 90°, implying a negative ac conductivity, which peaks at a drive frequency not far below the Bloch frequency.

## 1. Introduction: The problem

In a 1971 paper, Ktitorov, Simin, and Sindalovskii (KSS) [1] discussed the frequency dependence of the small-signal complex conductivity $\sigma(\omega)$ of a semiconductor superlattice under conditions of an electron relaxation time sufficiently long and a dc bias field $E_0$ sufficiently high that Bloch oscillations would play a key role in the electron dynamics. This was a first major theoretical step beyond the groundbreaking 1970 paper by Esaki and Tsu [2] that predicted the possibility of a negative differential conductivity (NDC) under those conditions, but which did not address the frequency dependence of the conductivity. In their work, KSS showed that the real part of the complex conductivity will initially become *more* negative with increasing frequency, until it reaches a resonance minimum at a frequency somewhat below the Bloch frequency, turning positive just below the Bloch frequency (Fig.1).

This negative-conductivity resonance close to the Bloch frequency makes a superlattice operating in this range an attractive gain medium for an active Bloch oscillator. In particular, it has been pointed out by Allen [3] that the resonance enhancement should make it possible to suppress any domain instabilities induced by a negative dc conductivity, by shunting the superlattice layers with a positive conductance that is sufficiently high to make the combined dc conductivity positive without obliterating the negative conductivity just below the Bloch frequency.

---

[*] kroemer@ece.ucsb.edu

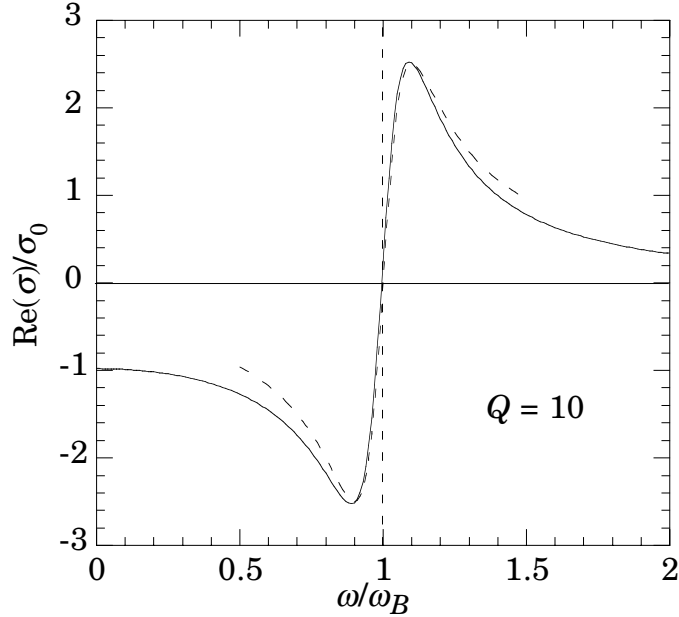

**Fig. 1.** Real part of the KSS differential conductivity, vs. frequency, assuming $Q = \omega_B \tau = 10$, where $\omega_B$ is the (angular) Bloch frequency and $\tau$ the electron relaxation time. See (3), (4), and (5) for the definitions of the quantities $\omega_B$, $\sigma_0$, and $Q$. The dashed line is a fit of the vicinity of the Bloch frequency to the approximation (40) derived later.

The principal objective of the present paper is to gain an in-depth understanding of the physical origin of this behavior, rather than having it "just come out of the math." Such a physical understanding becomes particularly desirable when one wishes to go beyond the small-signal conductivity treatment of KSS, and investigate the current dynamics under conditions that are likely to be important in the operation of future "real" Bloch oscillator devices, namely, when the ac drive field is no longer small compared to the dc bias field. It has been shown by Ignatov et al. [4, 5] that under such conditions the current dynamics can become remarkably complex, exhibiting features that go far beyond what might be expected from the KSS treatment.

In KSS, the authors considered a one-dimensional superlattice potential with period $a$, and with a simple electron dispersion relation of the form

$$\mathcal{E}(k) = \tfrac{1}{2}\mathcal{E}_0 \cdot (1 - \cos ka), \tag{1}$$

where $\mathcal{E}_0$ is the miniband width. They solved the Boltzmann transport equation in the relaxation time approximation. For simplicity, we ignore here their distinction between energy- and momentum relaxation times, and assume a common relaxation time $\tau$ for both processes. The KSS expression for the small-signal complex conductivity $\sigma$ may then be written in the form

$$\sigma(\omega) = \sigma_0 \frac{\left(\omega_B^2 \tau^2 - 1\right) + i\omega\tau}{\left[\left(\omega^2 - \omega_B^2\right)\tau^2 - 1\right] + 2i\omega\tau}. \tag{2}$$

Here,

$$\omega_B = \frac{1}{\hbar} e |E_0| a, \tag{3}$$

is the (angular) **Bloch frequency**, with which the electrons circulate through the Brillouin zone (BZ) of reduced $k$-space, and

$$\sigma_0 = \frac{\sigma_{00}}{Q^2 + 1}, \tag{4}$$

where $\sigma_{00}$ is the low-field dc conductivity, and

$$Q \equiv \omega_B \tau \tag{5}$$

is the central parameter characterizing the relative role of Bloch oscillations and scattering in the electron dynamics.

In the dc limit ($\omega = 0$), (2) reduces to

$$\sigma(0) = -\sigma_{00} \cdot \frac{Q^2 - 1}{\left(Q^2 + 1\right)^2}, \tag{6}$$

Note that $\sigma(0)$ is negative for $Q > 1$, which is the earlier result of Esaki and Tsu [2] (for a more elementary derivation, see Appendix A). Because of (6), the parameter $Q$ may also be expressed in the form

$$Q = E_0/E_c, \tag{7}$$

where $E_c$ is the *critical field* of the current density-vs.-field characteristic, defined as the field at which the static drift velocity peaks and the static differential conductivity goes through zero. Evidently, the desirable high-$Q$ operation requires a bias field large compared to the critical field [3].

Throughout the present paper, we will assume

$$Q \gg 1. \tag{8}$$

For sufficiently large values of $Q$, the resonance minimum occurs approximately at the frequency $(1 - 1/Q) \cdot \omega_B$, with a resonant enhancement factor of approximately $Q/4$ relative to $\sigma(0)$, as illustrated in Fig. 1 for $Q = 10$. The conductivity crosses over to positive values at the frequency

$$\omega = \omega_B \cdot \left(1 - 1/Q^2\right), \tag{9}$$

which for large values of $Q$ is just below $\omega_B$.

The behavior of the real part of $\sigma$ is itself the result of a resonant peak in the absolute magnitude of $\sigma$, almost exactly at the Bloch frequency $\omega_B$ (Fig. 2). However, at $\omega_B$, there is a phase shift very close to $\pi/2$ between current and driving field, hence the current is almost purely reactive. To obtain a negative differential conductivity, a phase shift larger than $\pi/2$ is required, and this is what happens at lower frequencies (Fig. 3). With decreasing frequency, the real part of the conductivity initially becomes more negative, but soon goes through a peak and becomes again less negative as $|\sigma|$ decreases away from its own resonance peak.

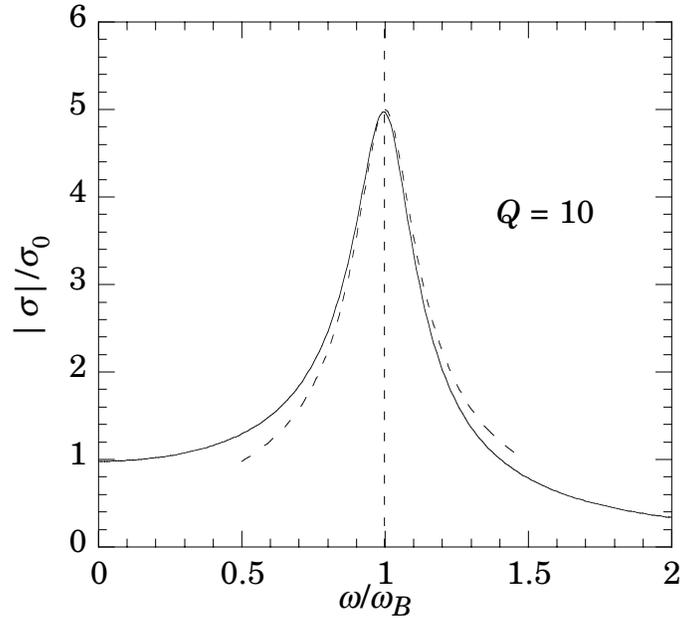

**Fig. 2.** Normalized absolute magnitude of the differential conductivity vs. normalized frequency, for the same parameters as in Fig. 1. The dashed line is a fit of the vicinity of the peak to the approximation (43) derived later.

For $Q \gg 1$, the peak absolute conductivity is given by

$$|\sigma|_{\max} \approx \tfrac{1}{2}\sigma_0 \cdot Q, \tag{10}$$

The peak in the real part of the negative conductivity is lower by another factor of 1/2, leaving the net enhancement by a factor $Q/4$ claimed above.

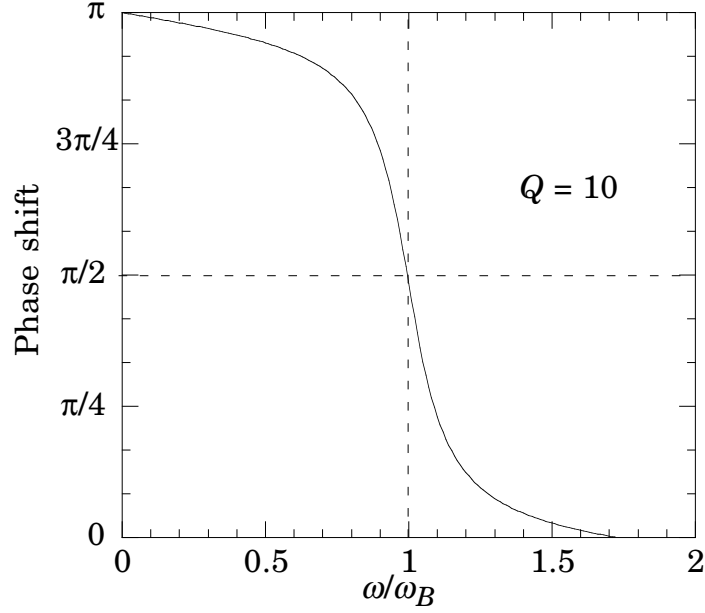

**Fig. 3.** Phase shift between the ac current density and the ac drive field, for the same parameters as in Figs. 1 and 2.

## 2. Electron bunching in *k*-space and negative conductivity

*2.1. External drive at the exact Bloch frequency*

In the presence of an applied force $F = -eE$, each electron propagates through (extended) $k$-space according to "Newton's law in $k$-space,"

$$\hbar \frac{dk}{dt} = F. \tag{11}$$

In a reduced-zone representation the electron undergoes an umklapp process whenever it reaches the zone boundary at $k = \pm\pi/a$, re-entering the reduced zone from the opposite boundary.

Given the dispersion relation (1), the associated electron velocity in real space is

$$v(k) = \frac{1}{\hbar} \cdot \frac{d\varepsilon}{dk} = v_{\max} \cdot \sin ka, \quad \text{where} \quad v_{\max} = \frac{1}{2\hbar} \cdot \varepsilon_0 a. \tag{12a,b}$$

Typical parameters for a superlattice might be $\varepsilon_0 = 30$ meV, $a = 20$ nm, leading to $v_{max} = 4.56 \times 10^7$ cm/s.

If only a dc force $F_0 = -eE_0$ acts on the electrons, the (extended) wave number $k$ of an electron increases linearly in time, and the velocity then oscillates purely sinusoidally with the Bloch frequency. The corresponding traversal time through the BZ is $t_B = 2\pi/\omega_B$.

In the limit of a an electron scattering frequency that is negligibly small compared to the Bloch frequency, the electrons will eventually become randomly distributed throughout the Brillouin zone, and their individual current contributions will average to zero. If we normalize the distribution function such that $f \cdot dk$ is the number of electrons in a $k$-space interval $dk$ wide, per unit superlattice length, the zero-order distribution function is simply a constant,

$$f_0 = N \cdot a / 2\pi, \tag{13}$$

where $N$ is the number of electrons per unit superlattice length. For the purpose of the present work, it will be useful to use this limit as the zero-order perturbation limit, rather than the thermal-equilibrium limit. Put differently, we view the dc bias field as part of the unperturbed problem, and the scattering as the perturbation.

In the presence of a non-zero scattering rate, a non-zero dc current will flow, which means that $f(k)$ acquires a component that is a odd function of $k$. We shall neglect this correction throughout the present paper: Not only does it decrease with increasing $Q$, but, more importantly, being a dc effect, it has only a very small effect on the topic of principal interest to us, the high-frequency resonant behavior near the Bloch frequency.

Assume now that a high-frequency field has been added to the dc field,

$$F(t) = F_0 + F_\omega \cdot \cos \omega t. \tag{14}$$

Integration of (11) leads to

$$k(t) = k(0) + \frac{1}{\hbar} F_0 t + \frac{1}{\hbar \omega} F_\omega \cdot \sin \omega t = k(0) + \frac{1}{a} \cdot \varphi(t), \tag{15}$$

where, in the last form, I have defined the quantity

$$\varphi(t) \equiv \omega_B t + \frac{F_\omega}{F_0} \cdot \sin \omega t, \tag{16}$$

which is the $k$-space phase of an electron that at $t = 0$ was located at $k = 0$. Throughout the present section we shall assume that the drive frequency $\omega$ is equal to the Bloch frequency $\omega_B$, deferring the more general case to section 2.3.

Consider now a distribution function of the specific form of a superposition of two traveling-wave components

$$f(k) = f_+(k) + f_-(k), \tag{17a}$$

where

$$f_\pm(k) = \frac{a}{2\pi} \cdot N_\pm \left\{ 1 \mp \cos[ka - \varphi(t)] \right\}, \tag{17b}$$

with, $N_- + N_+ = N$.

In the absence of scattering, each of the two distribution functions $f_\pm$ represents, by itself, a valid solution of the electron propagation problem through $k$-space, with every electron obeying Newton's law (11). Hence, in this limit, any superposition of terms of the form (17), with arbitrary values of $N_+$ and $N_-$ will also be a valid solution, provided both $N_+$ and $N_-$ are positive and add up to $N$. A special case would be $N_- = N_+ = N/2$, in which case the distribution function returns to the uniform limit

More general solutions would contain higher $k$-space harmonics, as well as sine-terms, but we ignore such terms here. We return to this point at the end of this section.

The central approximation made in the present paper is that we continue to use a distribution function of the form (17), with different occupancies $N_+$ and $N_-$, the ratio of which is determined by the scattering processes. I claim that the scattering causes an imbalance between the numbers of electrons in the two distribution functions, with $N_+ > N_-$.

To see this, note first that the addition of a high-frequency force at the exact Bloch frequency does not change the time $t_B$ required for an electron to traverse the BZ, it continues to be $t_B = 2\pi/\omega_B$. However, the speed of traversal through the BZ is now no longer uniform: During the "negative" half of the traversal time (i.e., while $F < F_0$), the speed is below the average speed $F_0/\hbar$, during the "positive" half ($F > F_0$) it is above. It is this speed difference that is ultimately responsible for the resonance of $|\sigma|$ displayed in Fig. 2, and for the negative-conductivity peak in Fig. 1.

Inspection of (14) and (17b) shows that the peak of the $f_+$ distribution passes through $k = \pm\pi/a$, the top of the miniband, at those instances of time at which the drive field goes through its maximum, $F_{\max} = F_0 + F_\omega$. But this is also the time at which the electron propagation velocity in $k$-space goes through its maximum. Now—and this is the central point of the argument—for any reasonable inelastic scattering process, the scattering probability must be higher from the upper half of the miniband to the lower half than the other way around, and it should be highest near the top of the miniband. The increased speed of propagation across the top therefore implies a *decreased* number of inelastic scattering events per transit, compared to the case of a pure-dc drive field.

For the $f_-$ distribution, the opposite holds: Its peak passes through the top of the miniband when the drive field, and hence the $k$-space velocity of the electrons, pass through their minima, which implies an *increased* number of inelastic scattering events.

The overall result is a transfer of electrons from $f_-$ to $f_+$. This transfer will continue until the population $N_+$ has build up (and $N_-$ has decreased) to such a level that the population difference between $f_+$ and $f_-$ balances the difference in scattering probabilities per electron. If the inelastic scattering rate is low, the number of transfers per transit will also be low, and it will require many transits to reach the final steady-state distribution. But, once reached, the final distribution will depend, to the first order, only on the *ratio* of the two scattering probabilities per transit.

If we assume that the approximation (17) remains applicable, the combination of (16) and (17) yields a distribution function

$$f(k) = f_0 \cdot \{1 - \eta \cdot \cos[ka - \varphi(t)]\}. \tag{18}$$

where

$$\eta \equiv \lim (N_+ - N_-)/N \tag{19}$$

is a parameter the value of which we will estimate in the next section.

Eq. (18) evidently represents a traveling wave in $k$-space superimposed on the unperturbed uniform distribution $f_0$. The (particle) current will then be given by

$$\begin{aligned} j(t) &= \int_{-\pi/a}^{+\pi/a} v(k) f(k) dk \\ &= -\frac{Na}{2\pi} \cdot v_{\max} \cdot \eta \cdot \int_{-\pi/a}^{+\pi/a} \sin ka \cdot \cos[ka - \varphi(t)] \cdot dk \\ &= -\tfrac{1}{2} N \cdot v_{\max} \cdot \eta \cdot \sin[\varphi(t)]. \end{aligned} \tag{20}$$

The electrical current density is obtained by multiplying with the electron charge –$e$; however, throughout this paper, all currents will be expressed as particle currents, and all fields in terms of the forces $F$.

The sine-term in (20) may be written as a Fourier-Bessel series (see Appendix B),

$$\sin[\varphi(t)] = \sum_{n=-\infty}^{+\infty} J_{n-1}(z) \cdot \sin[n\omega t], \tag{21}$$

where $J_n$ is the $n$th-order Bessel function, and

$$z = F_\omega / F_0. \tag{22}$$

Note that $J_{-n} = (-1)^n J_n$.

The overall current is not purely sinusoidal, but contains contributions at the various harmonics of the drive frequency, and a dc term. We are interested here only in the current density $j_\omega(t)$ at the fundamental frequency $\omega$,

$$j_\omega(t) = -I_\omega \cdot \sin(\omega t), \tag{23a}$$

with the amplitude

$$I_\omega = \tfrac{1}{2} N \cdot v_{\max} \cdot \eta \cdot [J_0(z) + J_2(z)] = \tfrac{1}{2} N \cdot v_{\max} \cdot \eta \cdot J_1(z)/z. \tag{23b}$$

In the last form, we have drawn on the Bessel function identity $J_0(z) + J_2(z) = J_1(z)/z$. There is no $\cos(\omega t)$-term.

In the limit $F_\omega \ll F_0$, of interest for comparing with KSS, we can expand the Bessel function,

$$J_1(z)/z \approx 1 - \tfrac{1}{8} z^2 \xrightarrow[z \to 0]{} 1. \tag{24}$$

Eq. (23b) then simplifies to

$$I_\omega = \tfrac{1}{2} N \cdot v_{\max} \cdot \eta. \tag{25}$$

However—and this point will become important later—the more general form (23) is by no means restricted to this limit.

The key aspect of the result (23) is that, in our treatment, the current is 90° out of phase from the ac drive field; it is the purely reactive current. It agrees with the KSS result in the limit $Q \gg 1$, with the correct amplitude. Note, however, that, for finite $Q$, it is only an approximation, albeit an excellent one—see Fig. 1 and Eq. (9). To reproduce the exact result, we would have had to include sine terms of the form $\pm\sin[ka-\varphi(t)]$ in (17a,b). A more detailed investigation [6] shows that these terms are small compared to the cosine terms, typically by a factor on the order $1/Q$, at least in the case considered here, when the drive frequency $\omega$ is equal to the Bloch frequency $\omega_B$.

The neglect of higher *k*-space harmonics in (17) is almost certainly an excellent approximation under the condition of interest to us: a relatively narrow miniband, and a scattering frequency small compared to the Bloch frequency.

*2.2. Simple scattering model*

Up to this point, we have made no assumptions about the scattering processes other than that they exist, and are associated with a positive energy dissipation, which in turn implies preferred scattering from the upper half of the miniband to the lower half. To estimate the parameter $\eta$, we must be more specific, and the value obtained will depend on the assumptions made.

It is probably reasonable to assume that the energy relaxation processes are strongest near the top of the miniband. It is then possible to obtain a simple estimate of $\eta$ by going to the—admittedly extreme—limit of assuming that all scattering takes place from a narrow interval of width $\Delta k << \pi/a$ centered about $k = \pm\pi/a$ at the top of the miniband, and neglecting all scattering from outside this interval. The instantaneous scattering rate at time $t$ for each of the two distributions is then proportional to the value of distribution function at $k = \pm\pi/a$,

$$f_\pm(\pi/a) = \frac{a}{2\pi} \cdot N_\pm \{1 \mp \cos[\pi - \varphi(t)]\} = \frac{a}{2\pi} \cdot N_\pm \{1 \pm \cos[\varphi(t)]\}. \tag{26}$$

The cosine can again be expressed as a Fourier-Bessel series (see Appendix B). Upon integration over one period, the oscillating terms in that series integrate to zero; the single non-oscillating term makes a contribution proportional to

$$\int_0^{t_B} \cos[\varphi(t)] dt = -\mathrm{J}_1(z) \cdot t_B, \tag{27}$$

with $z$ from (22).

In steady state, the scattering rates from the two distributions must cancel,

$$N_+ \cdot [1 - \mathrm{J}_1(z)] = N_- \cdot [1 + \mathrm{J}_1(z)], \tag{28}$$

which implies

$$N_\pm = \tfrac{1}{2} N \cdot (1 \pm \mathrm{J}_1(z)), \tag{29}$$

and the parameter $\eta$ reduces to

$$\eta = \mathrm{J}_1(z), \tag{30}$$

In the limit of a weak ac driving field, $F_\omega \ll F_0$, the Bessel function can again be expanded,

$$\mathrm{J}_1(z) \approx \tfrac{1}{2} z = \tfrac{1}{2} F_\omega / F_0 \,, \tag{31}$$

implying

$$\eta = \tfrac{1}{2} F_\omega / F_0 \,, \tag{32}$$

giving the expected linear dependence of the ac current density on the strength of ac drive force in that limit.

With increasing ac drive field, the parameter $\eta$ increases sub-linearly, and reaches a maximum of about 0.7 for $F_\omega \approx 1.8 F_0$. However, at the same time the Bessel function term in (23b) decreases; what matters is not $\mathrm{J}_1$ alone, but the product $\mathrm{J}_1^2 / z$, which goes through a broad maximum of 0.42 at $z \approx 1.36$, and declines towards zero beyond that.

*2.3. Negative conductivity below the Bloch frequency*

Assume now that the drive frequency is lowered below the Bloch frequency,

$$\omega = \omega_B - \delta\omega, \qquad (0 < \delta\omega \ll \omega_B). \tag{33}$$

The electron scattering will continue to lead to the formation of traveling electron bunches, in synchronism with the drive frequency. But, once generated, these bunches will oscillate through the BZ, not at the drive frequency, but at the higher Bloch frequency. Consider a bunch the center of which, at the time $t = t_0$, was located at $k = k_0$. After exactly one drive cycle, this bunch will not again be centered at $k_0$, but will have propagated past that point by an additional distance

$$\delta k \approx \frac{2\pi}{a} \cdot \frac{\delta\omega}{\omega_B}, \tag{34}$$

to a location $k_1 = k_0 + \delta k$, attenuated somewhat due to scattering out of the bunch. After $n$ drive cycles, it will have propagated to $k_1 = k_0 + n\delta k$, attenuated proportionately more.

Consider, more specifically, a bunch with $k_0 = -\pi/a$, at the top of the miniband, where it makes no contribution to the current. After one drive cycle, following an umklapp process, this bunch will be located to the right of $-\pi/a$, in a region where the group velocity is negative, opposite to the direction of the drive field. It is this phase shift that is ultimately responsible for the appearance of a negative differential conductivity.

The overall density wave present in $k$-space at any given instant of time, will be a superposition of individual bunches spread out in $k$-space, created during earlier drive cycles, with the older bunches being farther advanced in phase, but also more attenuated. The net result is a current oscillation that is attenuated in amplitude relative to the case $\omega = \omega_B$ and—more importantly—phase-shifted relative to the drive field by *more* than 90°. But this is exactly the behavior of Fig. 1: The phase shift implies a negative conductivity, initially increasing with increasing $\delta\omega$, but eventually the increasing attenuation implies a decrease of the magnitude of the negative conductivity to its dc value

If we ignore contributions from higher harmonics, and also ignore the small residual phase shift associated with a small but non-zero scattering rate, the overall current may be written as a superposition of terms of the form (23a),

$$j_\omega(t) = -i_\omega \cdot \sum_{n=0}^{\infty} \exp(-n t_D / \tau_a) \cdot \sin(\omega t + n \cdot \delta\omega \cdot t_D). \qquad (35)$$

Here, $t_D = 2\pi/\omega$ is the drive period, and $\tau_a$ is a relaxation time that determines the attenuation of each bunch with time. It is presumably related to the $\tau$ of the KSS theory, but not necessarily identical to the latter. The single-bunch current amplitude $i_\omega$ must be chosen such that in the limit $\delta\omega = 0$ we recover the result (23b). Evidently, for $\tau_a \gg t_D$,

$$I_\omega = \frac{i_\omega}{1 - \exp(-t_D / \tau_a)} \approx i_\omega \cdot \frac{\tau_a}{t_D} = i_\omega \cdot \frac{\omega \tau_a}{2\pi}, \qquad (36)$$

The total current in (35) may be split into a conductive ($\cos\omega t$) and reactive ($\sin\omega t$) component. We are principally interested in the former, which may be written as

$$j_{\sigma,c}(t) = -i_\omega \cdot \cos\omega t \cdot \sum_{n=0}^{\infty} \exp(-n t_D / \tau_a) \cdot \sin(n \cdot \delta\omega \cdot t_D). \qquad (37)$$

The sum can be reduced to a geometric series and evaluated in closed form. In the limits of interest to us, $t_D \ll \tau_a$ and $\delta\omega \ll \omega_B$,

$$\sum_{n=0}^{\infty} \ldots \approx \frac{1}{2\pi} \cdot \frac{\tau_a \delta\omega}{1 + (\tau_a \delta\omega)^2}. \qquad (38)$$

Insertion into (37) yields

$$j_{\sigma,c}(t) = -I_\omega \cdot \frac{\tau_a \delta\omega}{1 + (\tau_a \delta\omega)^2} \cdot \cos\omega t. \qquad (39)$$

which implies the real part of the conductivity

$$\text{Re}[\sigma(\omega)] = \frac{-I_\omega}{F_\omega} \cdot \frac{\tau_a \delta\omega}{1+(\tau_a \delta\omega)^2}. \tag{40}$$

Given the sign convention (33) for $\delta\omega$, we evidently have a negative conductivity for $\omega < \omega_B$, peaking at $\delta\omega = 1/\tau_a$, at which point

$$j_{\sigma,c}(t) = -\tfrac{1}{2}I_\omega \cdot \cos\omega t, \tag{41}$$

with an amplitude one-half that of the current density at $\omega_B$ in (23a).

By analogous arguments one obtains the imaginary part and the absolute magnitude of the conductivity:

$$\text{Im}[\sigma(\omega)] = -\frac{I_\omega}{F_\omega} \cdot \frac{1}{1+(\tau_a \delta\omega)^2}, \tag{42}$$

$$|\sigma(\omega)| = \frac{I_\omega}{F_\omega} \cdot \frac{1}{\sqrt{1+(\tau_a \delta\omega)^2}}. \tag{43}$$

As is shown by the broken lines in Figs. 1 and 2, in the vicinity of the Bloch frequency our approximations (40) and (43) are in excellent agreement with the more rigorous KSS results, provided we equate our attenuation time constant $\tau_a$ to the $\tau$ of the KSS theory. The two fits were obtained by assuming $\omega_B \tau_a = Q$ (= 10) and $I_\omega/F_\omega = \sigma_0 Q/2$.

*2.4. Sub-harmonic drive*

The idea of *k*-space bunching leads to some interesting predictions for the case that the drive frequency is an integer fraction of the Bloch frequency, or slightly below such a fraction.

Consider once again a bunch the center of which, at the time $t = t_0$, was located at $k = k_0$. If the drive frequency is exactly half the Bloch frequency, this bunch will have returned to $k = k_0$ after one full drive cycle, similar to drive at the Bloch frequency itself, but attenuated twice as much. As a result, we would expect a resonance under sub-harmonic drive, similar to the resonance for $\omega = \omega_B$, but significantly weaker, due to the reduction of the generation rate of the bunches, and the greater attenuation per bunch. If the drive frequency is slightly below one-half $\omega_B$, we would again expect a (weaker) negative real part of the conductivity, at least under otherwise favorable conditions. Similar but progressively weaker anomalies might be expected at higher-order sub-harmonics.

The KSS theory clearly does not show any special behavior at the sub-harmonics of the Bloch frequency. But exactly such anomalies were found for high ac drive amplitudes in the numerical work of Ignatov et al. [4], illustrating the predictive power of the *k*-space bunching concept.

From the point of view of a Bloch oscillator as a practical power-generating device, this sub-harmonic operation is probably not of interest. But measurements in the sub-harmonic range might be of diagnostic value in understanding the scattering processes.

## 3.  Discussion

We have shown that the negative-conductivity resonance can be understood as the result of the cooperation of two simple physical processes: (*i*) The ac component of the drive field induces a periodic modulation of the velocity with which the electrons traverse the BZ of $k$-space. (*ii*) In the presence of this velocity modulation, inelastic scattering events cause the formation of traveling electron bunches in $k$-space. It is these bunches that cause both the reactive current resonance in the immediate vicinity of $\omega_B$ and the negative conductivity at lower frequencies where the phase shift between drive field and current density exceeds 90°.

Although less quantitatively rigorous than the treatment in KSS, the very absence of some of the specific assumptions made in KSS give our model a greater generality, providing conceptual guidance for what to expect under conditions beyond those assumed in KSS. For example, the KSS treatment is a treatment only of the small-signal differential conductivity, while our $k$-space bunching concept carries no such restriction. In fact, it naturally lends itself to a treatment of what happens when the ac drive field is large, at which point both the velocity modulation and the resulting electron bunching are particularly strong. This point is important, because an understanding of large-amplitude effects is central to the understanding of the operation of hypothetical future Bloch oscillators, in at least two ways: saturation effects and domain instabilities. It is useful to comment briefly on both aspects.

If a negative-resistance element is inserted into a linear resonance circuit, the amplitude to which the oscillations will build up is determined by the non-linearities of the device.

Probably even more important is the problem of space-charge instabilities, which may occur in a medium with a *bulk* negative differential conductivity, *if* the latter extends down to zero frequency. Under bias, such a medium may break into spatial domains of different electric fields, similar to the well-known domains in the Gunn effect. These domains in turn tend to suppress the negative *overall* conductivity of the device at the intended oscillation frequency just below the Bloch frequency. However, this breakup may be suppressed *dynamically* if the ac drive signal is kept sufficiently large. For example, in the Gunn effect, a domain-free mode of large-amplitude operation exists, the so-called LSA mode (= **L**imited **S**pace charge **A**ccumulation mode; for an elementary review see [7]). There, the overall field dips, during each cycle, to very low values at which the *static* velocity-field-characteristic has a steep positive slope. Under steady-state operation at such low fields, the medium would be an "ordinary" conductor with a positive conductivity, and

4any space charges would decay rather than build up. Under suitable ac operating conditions, domains will then be unable to build up. One might suspect that a similar domain suppression might be achievable in a Bloch oscillator, despite the significantly different origin of the negative conductivity in the two phenomena.

A quantitative study of Bloch oscillations under large-signal drive conditions, based on a more rigorous elaboration of the *k*-space bunching concept, has confirmed that expectation [6]. However, the details are significantly different from those in the Gunn-LSA mode. A discussion of this topic would go beyond the scope of the present paper; it will be presented in a separate paper.

The most important limitation of our treatment—which it shares with KSS and most other treatments in the literature—is probably not our set of simplifying assumptions about the distribution function, but our restriction to a one-dimensional dynamics. In layer-type superlattices, the electron motion parallel to the layers is not quantized, and is associated with the additional kinetic energy

$$\mathcal{E}_{xy} = \frac{\hbar^2}{2m^*} k_{xy}^2, \tag{44}$$

where $k_{xy}$ is the wave number in the *xy*-plane. If we designate the "1-D energy" given in (1) as $\mathcal{E}_z$, we have a total energy of the form

$$\mathcal{E} = \mathcal{E}_z + \mathcal{E}_{xy}, \tag{45}$$

with no true upper band edge for the miniband. If we plot the energy as a function of $k_{xy}$, we obtain a diagram as in Fig. 4, which shows two parabolas representing the top and the bottom of the 1-D miniband as functions of $k_{xy}$. By a combination of Bloch oscillations and elastic scattering events an electron can now acquire energies much larger than the 1-D miniband width $\mathcal{E}_0$. In particular, highly inelastic optical phonon scattering can now take place even if the 1-D miniband width $\mathcal{E}_0$ is less than the optical-phonon energy $\hbar\omega_{op}$, a sequence illustrated in Fig. 4.

What must be avoided is the possibility of electron transfer into higher minibands, which calls for minigaps wide compared to the optical phonon energy.

Obviously, an inclusion of these 3-D processes greatly increases the complexity of the mathematical treatment, calling for numerical methods if *quantitative* results are to be obtained, for example by an extension of the 1973 Monte Carlo calculations modeling of Anderson and Aas [8]—a task outside the scope of the present paper.

But none of this invalidates the basic concept of *k*-space electron bunching as the mechanism underlying the negative conductivity. Even the concept of a 1-D distribution function retains much of its validity; it may be viewed as the 3-D function integrated over the in-plane $k$-space. In fact, the increase in mathematical complexity required for a quantitative solution of the dynamics problem probably enhances the need for a simple conceptual guiding principle like the $k$-space bunching concept, rather than diminishing it.

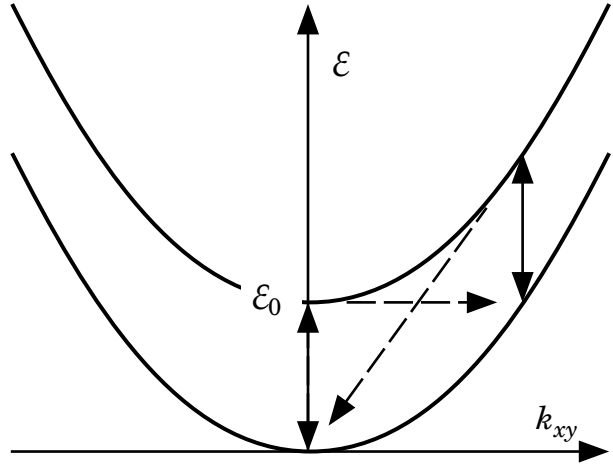

**Fig. 4.** Cross-sectional miniband diagram in the presence of un-quantized motion parallel to the superlattice layers, and a potential scattering sequence in such a structure. The two parabolas represent the top and the bottom of the 1-D miniband as functions of the in-plane wave number $k_{xy}$ Bloch oscillations (vertical double arrows), combined with elastic scattering (horizontal arrow), permit an energy buildup beyond the 1-D miniband width, and inelastic scattering (oblique arrow) may take place even if the 1-D miniband width $\mathcal{E}_0$ is less than the optical-phonon energy.

## Acknowledgments

I wish to thank Prof. S. J. Allen for bringing the KSS paper to my attention, and for numerous useful discussions.

## Appendix A: Negative differential dc conductivity

The negative differential dc conductivity for large values of $Q$ is easily understood in terms of a little-known argument dating back to 1953 [9], which is more elementary than the treatment of Esaki and Tsu [2],

Consider an electron oscillating back and forth in a tilted band of finite width, as illustrated in Fig. 5. If no energy loss events would take place, there could be no dc current flowing (a dc current in the presence of a dc field requires a dissipation mechanism), and the oscillatory contributions of the different uncorrelated electrons would average to zero.

If now the electron loses an energy $\Delta\mathcal{E}$ due to, say, an optical-phonon scattering event, the oscillation will be displaced downstream by $\Delta x = \Delta\mathcal{E}/eE$, and if such scattering events occur with a frequency $v = 1/\Delta t$, the electron travels with a drift velocity

$$v = \overline{\left(\frac{\Delta x}{\Delta t}\right)} = \frac{1}{eE} \cdot \overline{\left(\frac{\Delta\mathcal{E}}{\Delta t}\right)}. \tag{46}$$

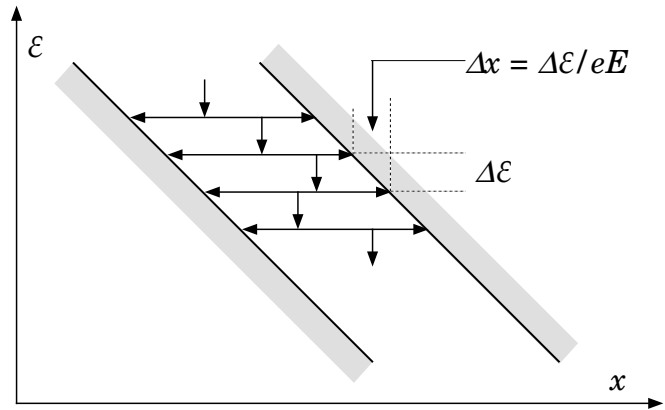

**Fig. 5.** Bloch oscillations combined with inelastic scattering as a mechanism for a decrease of drift velocity with increasing bias field $E$ (adapted from ref. [9]).

So far, this is simply the law of conservation of energy. But once the field has reached such a magnitude that several Bloch cycles pass between successive scattering events, then both the collision frequency and the energy loss per collision will saturate, and the net result is a decrease of drift velocity with increasing field, roughly inversely proportional to the field.

At small values of the electric field, the energy loss per collision will drop. In fact, in the limit of zero field, there can be no average energy loss per collision at all. For small values of $E$, $\Delta\mathcal{E}$ must be an even function of $E$, starting proportional to $E^2$. This leads to a drift velocity linear in $E$—as it must.

The negative-conductivity consequence of this behavior was not pursued in [9]; it was subsequently pointed out by Shockley and Mason [10] that the implied negative differential conductivity might be utilizeable for high-frequency amplification and signal generation up to frequencies beyond those achievable with three-terminal devices. Unfortunately, these ideas proved unworkable in bulk semiconductors, where the short lattice period implied electric fields far above those at which avalanche breakdown sets in. But with the much longer periods of artificial superlattices, these ideas moved into a regime where such oscillators might become realizable, and it appears worthwhile to resurrect the above elementary argument.

It can be shown that (46) agrees quantitatively with the high-$Q$ limit of the KSS result (6) if one makes the identification

$$\frac{\Delta \mathcal{E}}{\Delta t} = \frac{\mathcal{E}_0}{2\tau_e}, \tag{47}$$

where $\tau_e$ is the energy relaxation time of the complete KSS theory, that is, without equating energy and momentum relaxation times as we have done here.

## Appendix B: Mathematical Detail

To evaluate (20) and (27), we draw on the familiar Bessel function relations (see [11])

$$\sin(z \cdot \sin\theta) = \sum_{n=-\infty}^{+\infty} J_n(z) \cdot \sin nz \quad \text{(odd } n \text{ only)}, \tag{48a}$$

$$\cos(z \cdot \sin\theta) = \sum_{n=-\infty}^{+\infty} J_n(z) \cdot \cos nz \quad \text{(even } n \text{ only)}, \tag{48b}$$

where $J_n$ is the $n$th-order Bessel function, and $J_{-n} = (-1)^n J_n$. From these, one easily derives the more general relations

$$\sin[\gamma + z \cdot \sin\theta] = \sum_{n=-\infty}^{+\infty} (-1)^n J_n(z) \cdot \sin(\gamma - n\theta), \tag{49a}$$

$$\cos[\gamma + z \cdot \sin\theta] = \sum_{n=-\infty}^{+\infty} (-1)^n J_n(z) \cdot \cos(\gamma - n\theta). \tag{49b}$$

In both (20) and (27), we have $\gamma = \theta = \omega_B t$ and $z = F_\omega/F_0$. Inserting these into (49a) leads to the claimed result (21). In (49b), we obtain

$$\cos[\omega_B t + z \cdot \sin \omega_B t] = \sum_{n=-\infty}^{+\infty} (-1)^n J_n(z) \cdot \cos[(1-n) \cdot \omega_B t]. \tag{50}$$

The non-oscillating $n = 1$ term evidently has the value $-J_1(z)$ claimed in (27).